\documentclass[%
 reprint,
 amsmath,amssymb,
 aps, pre
]{revtex4-1}

\usepackage{graphicx}
\usepackage{dcolumn}
\usepackage{bm}
\usepackage{hyperref}
\usepackage[mathlines]{lineno}

\hypersetup{
    colorlinks=true,
    linkcolor=blue,
    filecolor=magenta,      
    urlcolor=cyan,
}

\usepackage{caption}
\usepackage{subcaption}

\begin{document}


\title{Intruder in a two-dimensional granular system: statics and dynamics of force networks in an experimental system experiencing stick-slip dynamics}

\date{\today}

\author{Rituparna Basak}
\affiliation{%
 Department of Mathematical Sciences, New Jersey Institute of Technology, Newark, New Jersey 07102, USA
}   

\author{Ryan Kozlowski}
\affiliation{College of the Holly Cross, Worcester, Massachusetts  01610, USA }

\author{Luis A. Pugnaloni} 
\affiliation{Departamento de F\'isica, Facultad de Ciencias Exactas y Naturales, Universidad Nacional de La Pampa, CONICET, Uruguay 151, 6300 Santa Rosa (La Pampa), Argentina}

\author{M.~Kramar}
\affiliation{Department of Mathematics, University of Oklahoma, Norman, Oklahoma 73019, USA}

\author{Joshua E.~S.~Socolar}
\affiliation{Department of Physics, Duke University, Durham, North Carolina 27708, USA}

\author{C. Manuel Carlevaro}
\affiliation{Instituto de F\'isica de L\'iquidos y Sistemas Biol\'ogicos, CONICET, 59 789, 1900 La Plata, Argentina \\
and Departamento de Ingenier\'ia  Mec\'anica, Universidad Tecnol\'ogica Nacional, Facultad Regional La Plata, Av.\ 60 Esquina 124, La Plata, 1900, Argentina}

\author{Lou Kondic}
\email{kondic@njit.edu}
\affiliation{%
 Department of Mathematical Sciences and Center for Applied Mathematics and Statistics, New Jersey Institute of Technology, Newark, New Jersey 07102, USA
}

\begin{abstract}
In quasi-two-dimensional experiments with photoelastic particles confined to an annular region, an intruder constrained to move in a circular path halfway between the annular walls experiences stick-slip dynamics. We discuss the response of the granular medium to the driven intruder, focusing on the evolution of the force network during sticking periods.  Because the available experimental data does not include precise information about individual contact forces, we use an approach developed in our previous work (Basak et al, J. Eng. Mechanics (2021)) based on networks constructed from measurements of the integrated strain magnitude on each particle.  These networks are analyzed using topological measures based on persistence diagrams, revealing that force networks evolve smoothly but in a nontrivial manner throughout each sticking period, even though the intruder and granular particles are stationary. Characteristic features of persistence diagrams show identifiable changes as a slip is approaching, indicating the existence of slip precursors.  Key features of the dynamics are similar for granular materials composed of disks or pentagons, but some details are consistently different. In particular, we find significantly larger fluctuations of the measures computed based on persistence diagrams, and therefore of the underlying networks, for systems of pentagonal particles.
 \end{abstract}

\maketitle

\section{Introduction}
\label{sec:introduction}

Stick-slip motion, characterized by intermittent yielding of a system subject to steadily increasing applied shear stresses, occurs in a plethora of systems, from metallic glasses to tectonic plates and has been explored extensively; see \cite{arcangelis_physrep_2016,falk_langer_2011} for reviews.  During the last two decades, significant attention has been paid to the possible connections between the microstructure of a system experiencing stick-slip dynamics and the global intermittent  behavior.  Studies of granular matter have considered the statics and dynamics of individual particles, including the possible existence of slip precursors that might inform predictions of upcoming large events~\cite{Talamali2011_pre,scuderi_natgeoscience_2016,lherminier_PRL_2019,johnson_geophys_2013,johnson_grl_2017,nerone_pre_2003,aguirre_pre_2006,scheller2006precursors,zaitsev_epl_2008,staron2002preavalanche,Maloney2004_prl,ciamarra_prl10,welker2011precursors,dahmen_natcom_2016,dahmen_group15,Denisov2017_sr,bares_pre17}. 

Stress fields in granular materials are generally characterized by the concentration of stress on chains of particles known as ``force chains.''   
In this paper, we follow a more general approach and instead of force chains consider force network as a whole.  
We show that topological  measures  based on persistence diagrams describing the force networks reveal nontrivial dynamics even when the intruder remains essentially stationary as the external force on it is increased during a sticking period.  We analyze experimental data by first extracting force networks of from each experimental image, processing them to obtain persistence measures, and then considering how those measures change during sticking events.
 
In the last few decades, a significant body of work on forces measured in experiments on granular materials has been conducted using photoelastic particles~\cite{daniels_hayman_2008, Hayman2011_pag,clark_prl12,tordesillas_bob_pre12,zadeh2019crackling,zadeh2019enlightening,kozlowski_pre_2019}.  Photoelastic particles' optical properties depend on the strains induced by applied boundary stresses, and high-resolution images of the strain fields in individual particles can be used to extract individual contact forces.  One technical but important issue, however, is that the resolution required for direct inference of the contact forces is not easily achieved.  In addition, inference of a force and torque balanced network of contact forces requires solving a computationally demanding inverse problem, so in practice, it is usually employed only for rather small systems.  Finally, the applicability of such methods has been largely limited to circular particles, although strides have been made recently for ellipsoidal ones~\cite{ellipsephotoelasticsolver_2021}. 

There are, however, other approaches to the quantification of stresses at the particle scale. A commonly implemented one is the so-called ``G$^2$'' method.  Forces applied to a photoelastic particle lead to the formation of intensity fringes when the particle is viewed through crossed polarizers.  It turns out that the sum of the normal force magnitudes on a particle is approximately proportional to the integrated gradient squared of the observed intensity signal. (See Ref.~\cite{daniels2017photoelastic} for a review of photoelasticity.)  This approach applies even to non-circular particles \cite{Geng2003greensfunction,YiqiuSingleG2} and is much less computationally demanding than solving for individual contact forces.  The result is an assignment of a stress magnitude to each particle, which preserves key features of the contact force data~\cite{basak_jem_2021}.  An even simpler approach might be to measure correlations and topological features directly from photoelastic images of the full system, without even attempting to identify individual particles or stresses.   

In previous works, we carried out a direct analysis of photoelastic images  to provide a basic understanding of the structure of force networks in the systems exposed to impact~\cite{pre18_impact} or shear~\cite{cheng_pg_2021}. Though these studies provided useful insights, they suffered from the limitations imposed by low experimental resolution and associated artifacts.  Motivated by this observation, we have started exploring the possibility that force networks constructed from G$^2$ information can yield similar quantitative and qualitative insights to networks constructed from exactly known particle-particle contact forces.  In our recent work,  Ref.~\cite{basak_jem_2021}, we made progress by showing that statistical measures of networks produced from simulation data with either interparticle contact forces (FC) or the total normal force magnitude on each particle (FP) yield consistent results (if the normal force on a particle $j$ due to contact with another particle $i$ is $\vec F_i$, then FP$_j= \sum_i |\vec F_i|$, where the sum goes over all contacts of particle $j$). In the present work, we construct these FP networks from experimental gradient (G$^2$) data and analyze their evolution.  We find that certain measures derived from persistence diagrams exhibit trends that indicate nontrivial restructuring of the force network during a sticking event and may aid in identifying signals of impending slip events.   

The rest of this paper is structured as follows. In Sec.~\ref{sec:methods}, we discuss the methods: first experimental ones (where more detailed information can be found in previously published work focusing on the same system~\cite{KozlowskiBasalFrictionPRE,KozlowskiCarlevaroSimulationBasalFrictionPoint,pugnaloni_pre_2022}); then the image processing methods used to extract FP networks, and finally the persistent homology-based methods that we employ to quantify the static and dynamic properties of the observed networks.   In Sec.~\ref{sec:results}, we describe the application of these methods to experimental data obtained from an intruder moving in an intermittent fashion through an annular domain containing photoelastic particles; first disks and then pentagons. Section~\ref{sec:conclusions} is devoted to the summary and conclusions.

\section{Methods}
\label{sec:methods}

\subsection{Experiments}\label{expdetails}

\begin{figure}[t!]
    \centering
    \includegraphics[scale = 0.3]{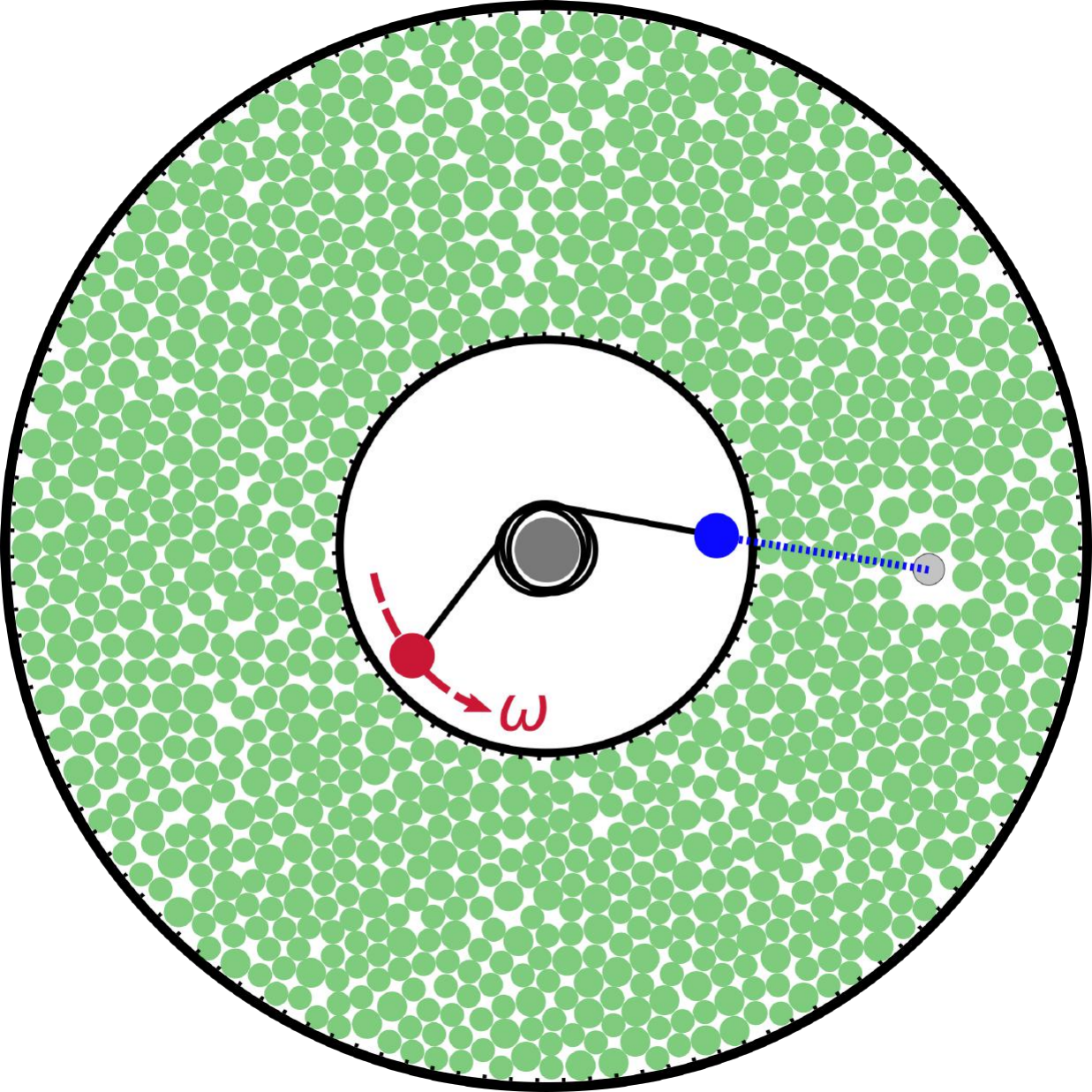}
    \caption{
    Schematic of the experimental setup.  An intruder is attached to a torsion spring (shown in blue), connected to a stepper motor rotating at a fixed rate $\omega$, that provides the driving force.  Granular particles are shown in green.
    } 
    \label{fig:exp}
\end{figure}

A bidisperse mixture of quasi-two-dimensional (2D) disks or regular pentagons compose the granular medium through which an intruding particle is pushed by a spring; see Fig.~\ref{fig:exp}. The 
number ratio of large to small particles is approximately 10:11 for both the disk-packing and the pentagon-packing. The diameters of the small and large disks, respectively, are $d_{\rm Ds} = 1.280 \pm 0.003$ cm and $d_{\rm Dl} = 1.602 \pm 0.003$ cm. The diameters of the circumscribing circles of the small and large regular pentagons, respectively, are $d_{\rm Ps} = 1.211 \pm 0.001$ cm and $d_{\rm Pl} = 1.605 \pm 0.001$ cm (close to the radii of the disks). The interparticle contact friction coefficient for all particles is $\mu = 0.7 \pm 0.1$, and the basal friction coefficient between particles and the glass base of the annulus (see below) is $\mu_{\rm BF} = 0.25 \pm 0.05$. 
We note that the typical horizontal forces applied to the intruder are on the order of Newtons, whereas the weight of a particle is on the order of $0.01\,$N.  This suggests that basal friction does not play an important role in the dynamics of strongly stressed particles, though it may be important for particles that lie outside the dominant force network at any given time.

The granular medium is confined to a fixed-volume annular cell with a radial width $W\approx 14d_{\rm Ds}$ between the inner and outer circular boundaries. Each boundary has a layer of ribbed rubber that prevents particles from slipping at the walls. The intruding particle, a cylinder with diameter $D = d_{\rm Dl}$, is suspended from a cantilever into the quasi-2D annular channel at a fixed radius halfway between the annulus walls.  One end of a torque spring ($\kappa = 0.4$ Nm/rad) in the central axis of the system is rotated at a low, fixed-rate $\omega = 0.12$ rad/s and the other end is coupled to the intruder. Slowly driving the spring-loaded intruder through the medium at high packing fractions results in stick-slip dynamics~\cite{kozlowski_pre_2019}; the medium exhibits \textit{sticking periods} of quasi-static \cite{gdr_midi_04} stress increase interspersed with intermittent, stress relieving \textit{slip events}. During these relatively rapid slip events, the medium plastically deforms and the intruder slides until a stable configuration of particles stops it, initiating a new sticking period~\cite{nasuno98,albert98bb}. 

Load cells in the cantilever that holds the intruder measure forces corresponding to the force exerted by the granular medium on the intruder in the azimuthal direction. Two cameras above the system visualize intruder and particle dynamics (via standard unfiltered photography) and photoelastic stresses within the granular medium (via dark-field polariscope imaging). We start collecting data, discussed further below, after the intruder has already traveled four full revolutions around the annulus, ensuring that we are observing statistical steady states. The images are recorded at the rate of 50 images/second.

Further details of the experimental data acquisition, along with analyses of intruder and particle dynamics, can be found in Ref.~\cite{kozlowski_pentagonsanddisks2021}.   The interested reader is also directed to Ref.~\cite{pugnaloni_pre_2022} for a discussion of the dependence of the results on system size.

\subsection{Image Processing}
\label{method_imgProcess} 
We record white-light (particle tracking) and photoelastic (stress-measurement) images in the experiments as described in Sec.~\ref{expdetails} and use these images to quantify force networks. The outline of our approach is as follows: We first find the average gradient-square (G$^2$) of the intensity of transmitted light for each particle. Then we convert G$^2$ for each particle to the magnitude of the normal force acting on each particle using independent calibration results (as mentioned in the introduction, while in principle G$^2$ measures particle strain, which could be then related to particle stress, it has been shown that this measure describes well the sum of the magnitudes of the normal forces acting on a particle, and this is what we are using here~\cite{YiqiuSingleG2}).  
Lastly, we create a force network. Specifically, for each experimental image, we use the following image-analysis protocol: 

\begin{figure}[t!]
    \begin{center}
         \includegraphics[width=.2\textwidth]{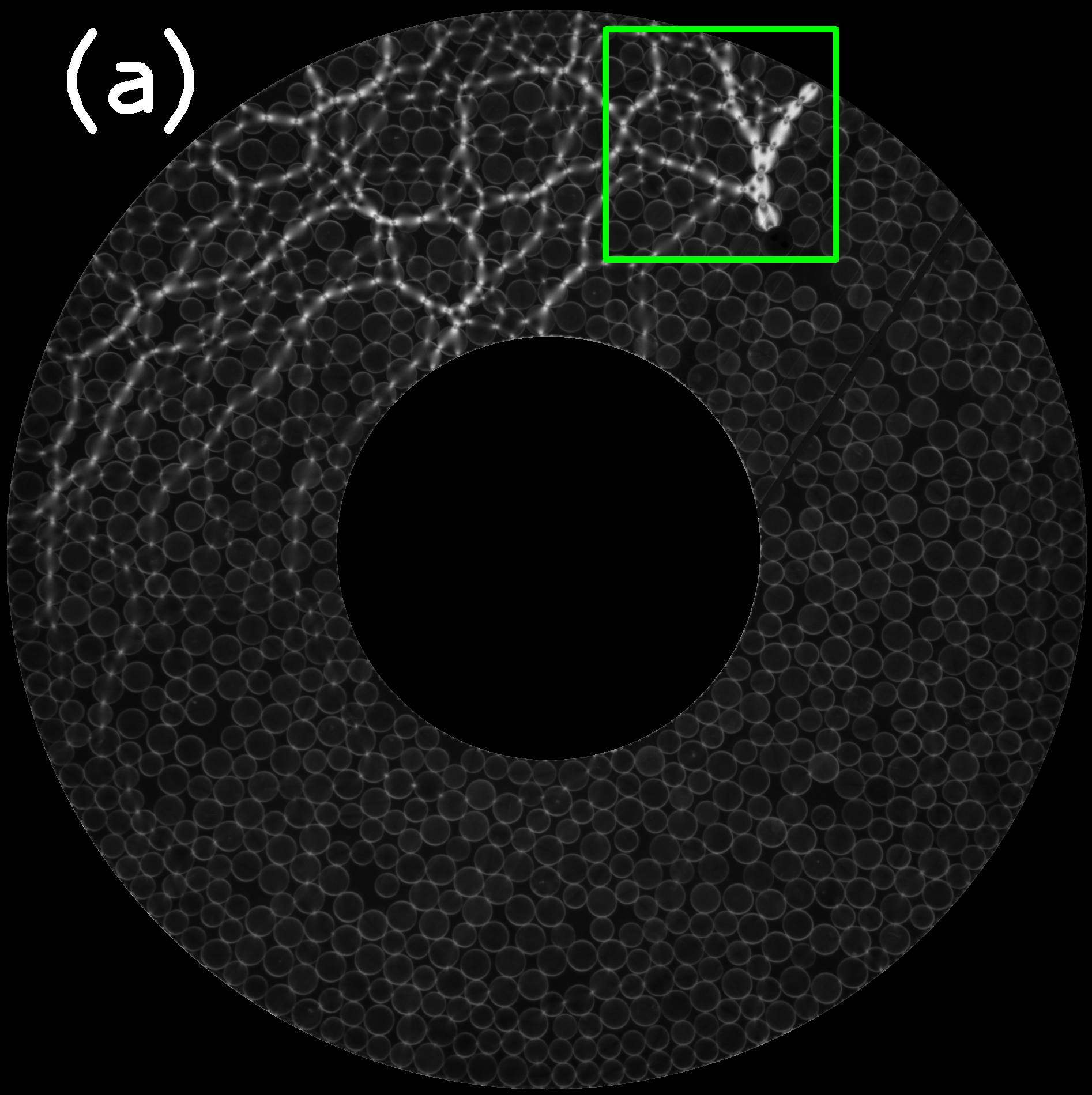}
         \includegraphics[width=.2\textwidth]{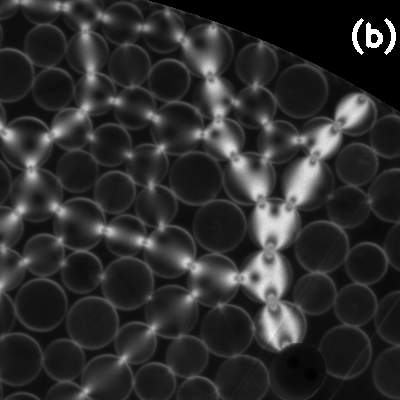}
          \includegraphics[width=.2\textwidth]{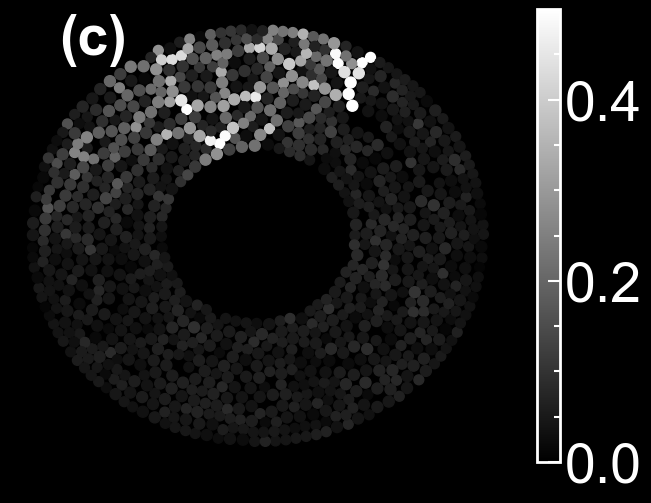}
          \includegraphics[width=.2\textwidth]{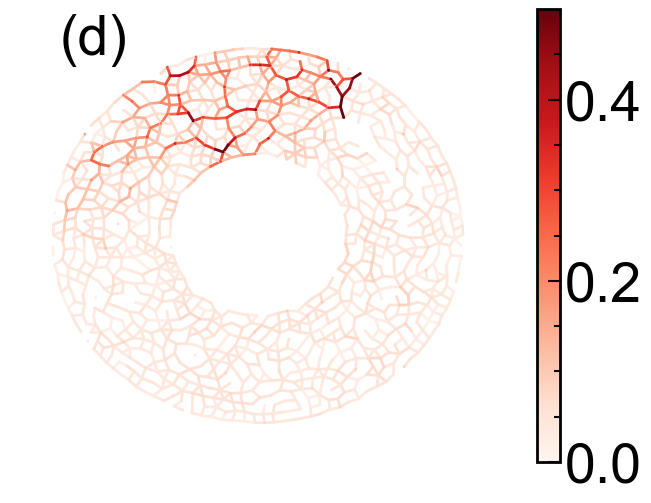}
    \end{center}
    \caption{
    (a) A photoelastic image of the annulus packed with disks. (b) Enlargement of the outlined region of (a) showing the details of the photoelastic signal. (c)  Processed image of the configuration from (a) showing force magnitude per particle. (d) The resulting force network. The color bars in  (c) and (d) show the forces normalized by the current value of the intruder force. (Note that some particles, in particular the ones close to the intruder, exceed the range shown on the color bar; we choose a maximum of 0.5 for visualization purposes.)  
}
          \label{fig:example_image}
      \end{figure}

$\bullet$ The squares of the intensity gradient values are calculated at each pixel location of a given photoelastic image. (See Fig.~\ref{fig:example_image}(a) and animations~\cite{sup_mat1}.) To identify the pixels inside a particle, the centers and orientations of tracked particles are found from the corresponding white-light image. The gradient-squared values of the pixels inside the particles are then averaged over the area of the particle, resulting in the average G$^2$ per particle. To avoid the experimental errors due to diffraction from particle edges, we decrease the area of selected pixels by 5$\%$, removing a band of pixels from a particle's boundary before calculating the average G$^2$.  For small particles, the width of the band is about 2 pixels. 

\begin{figure}[t!]
    
         \includegraphics[width=.35\textwidth]{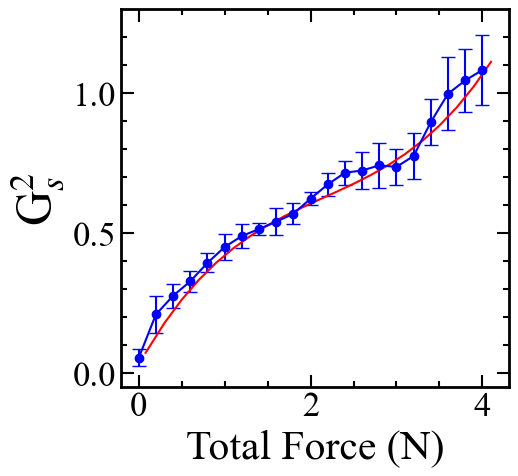}
    \caption{The G$_s^2$ data from a single disk experiment while pushing a disk against the intruder, which is held in place. The horizontal axis shows the total force applied on a particle, which is twice the applied force in these experiments.  Error bars show the standard deviation, and the calibration curve (in red) is the best cubic fit to the data.   
}
             \label{fig:calibration curve}
\end{figure}

$\bullet$ The G$^2$ per particle is then converted to the force magnitude measured in Newtons using a calibration curve obtained from an independent single disk experiment in which the disk is biaxially loaded by a controlled force; the calibration curve is shown in Fig.~\ref{fig:calibration curve}. To convert G$^2$ per particle to the force magnitude, we use a third-degree polynomial fit. The considered force range covers the forces experienced by photoelastic particles in our main experiments. 
Figure~\ref{fig:example_image}(c) and~\cite{sup_mat1} (movie 2) illustrate the final result.  Some aspects of these results, in particular the ones related to the scaling of the sample-averaged G$^2$ signal with the intruder force, are discussed below in Sec.~\ref{sec:example}. We note that the use of the calibration curve implicitly assumes that the G$^2$ information is independent of the number of contacts, which for our main experiments is typically more than two.

$\bullet$ We determine whether disks with radii $r_i$ and $r_j$ are in contact by checking whether the distance between the centers of the two disks is less than $1.05(r_i + r_j)$, accounting for experimental uncertainties in the tracked particle positions. We have confirmed that adding 5\% of the particle radius is sufficient for ensuring that very few contacts are missed (estimated less than $0.1\%$), while the added distance is sufficiently small so that artificial contacts are not produced. In the case of pentagons, we determine contacts by detecting overlaps between each pair of particles using the SHAPELY package~\cite{gillies2013shapely}. From the tracked particle positions and orientations, we input the particle vertices into SHAPELY after scaling the pentagon apothem by $1.05$. Again, we have checked that this increase in particle size allows the detection of $\approx 99.9\%$ of contacts and does not produce artificial ones.

$\bullet$ To construct a force network, we assign to each particle the force magnitude obtained from the G$^2$ calibration curve, then, for each pair of particles $p_i$ and $p_j$ in contact, assign to the edge $v_{i,j}$ joining their centers the minimum of the two force magnitudes, following the approach described in Ref.~\cite{basak_jem_2021}.  Note that within the presently implemented approach, the nodes of the networks are the particle centers and the edges are the lines connecting centers of the particles in contact.    Figure~\ref{fig:example_image}(d) shows a force network obtained by this procedure; see also~\cite{sup_mat1} (movie 3) for an animation. Parts (c) and (d) of Fig.~\ref{fig:example_image} are normalized by the current value of the intruder force, as discussed in more detail in Sec.~\ref{sec:results}.   

      \subsection{Computational topology}

The outcome of the approach described in the preceding section is a series of force networks that we analyze using tools based on persistent homology, which is a discipline of computational topology, see~\cite{mischaikow}.   In particular, we obtain persistence diagrams (PDs) that encode the main features of the underlying networks while at the same time providing significant data reduction. Furthermore, and perhaps most importantly in the present context, once a contact network is given, PDs are known to be stable with respect to small perturbations of the input data.  We refer the reader to our earlier works~\cite{basak_jem_2021,physicaD14} describing PDs and their properties in some detail. Here we provide only the outline of the approach and of the computations that are carried out.  

The PDs considered in the present work are computed by considering the structure of the force network's  structure exceeding any given force threshold and keeping track of the appearance (birth) and disappearance (death) of features (parts of the network) as we decrease the threshold. A PD is a collection of points with coordinates that encode birth and death thresholds of the components (or loops) present in the force network. In two spatial dimensions, the topological structures of interest are connected components (roughly corresponding to `force chains' in the case of strong forces) and loops/cycles.   All PDs reported in the present work are computed using the GUDHI library~\cite{gudhi}.

Given the computed PDs, we proceed by considering two separate sets of measures.  One of them, total persistence, TP, is simply the sum of the distances of the points (generators) in the PDs from the diagonal.  A landscape analogy may help clarify the physical meaning of TP: each generator corresponds to a peak in the force landscape, and the TP measures the sum over generators of the height of the peak above the highest saddle point connecting it to a higher peak.  Note that there is always one generator on the horizontal axis, corresponding to the highest peak in the network. Similar but more elaborate analogies could be produced for the case of loops.   TP accounts for the fact that the points that are further away from the diagonal carry more importance; the points close to the diagonal have similar birth and death coordinates, meaning that they {\it persist} only over a small range of force values, and therefore may not be significant.  This fact is particularly  relevant for the experimental data considered in the present work, which include some degree of experimental inaccuracy.  Furthermore, experimental limitations motivate us to exclude from consideration the generators that are very close to the diagonal; more precisely, we will not consider the generators whose distance to the diagonal is less than 0.1N, the estimated precision of the photoelastic analysis of the force magnitude  on a particle in our experiments. 

The other measure we use quantifies differences (distances) between two PDs.  The computations are again carried out using the GUDHI library~\cite{gudhi}. The distance between two PDs is computed as the minimum of the sum of all distances by which the points from one PD need to be translated to match the ones in the other PD.  If the diagrams have different numbers of points, then the extra points are mapped (vertically) to the diagonal. Various norms could be used for computing this distance, and in the present work, we consider two: the Wasserstein distance, W$_2$, which is essentially the $L_2$ norm, and the bottleneck distance, which is the $L_\infty$ norm.  See~\cite{kramar_physD_2016} for more details.  

\section{Results}
\label{sec:results}

We are now ready to discuss the results of our analysis of force networks.  To begin, in Sec.~\ref{sec:example} we present time series results that illustrate a close correlation between intruder dynamics and evolving networks.  Similarly to the rest of the discussion in this section, we focus on the stick phase of the intruder dynamics and in particular on the evolution of force networks during periods when the granular system is essentially stationary (but the force on the intruder is progressively increasing).  We then discuss time-averaged results, with the goal of reaching a general understanding of the force network's statics and dynamics during large numbers (hundreds) of stick periods. Results for disks  are presented in Sec.~\ref{sec:disks}.  The results for pentagons are given in Sec.~\ref{sec:time_dep_pentagons} (time series), and Sec.~\ref{sec:pentagons} (time-averaged results). For disks, we mostly focus on the particle volume fraction of $\phi = 0.77$; the discussion refers to this volume fraction unless otherwise specified. In addition, we show more limited results for $\phi = 0.75$ and $\phi = 0.72$.  For pentagons, the volume fractions considered are $\phi = 0.65$ and $\phi = 0.61$;  $\phi=0.65$ is assumed if not specified otherwise.  The largest values of $\phi$ for each particle geometry (disks and pentagons) are close to the maximum possible one in the considered experiments; for even slightly larger $\phi$, there is a significant likelihood of out-of-plane buckling during stick periods.

\subsection{Examples of time-dependent results for disks}
\label{sec:example}        

\begin{figure}[t!]
\centering
\includegraphics[width =.45 \textwidth]{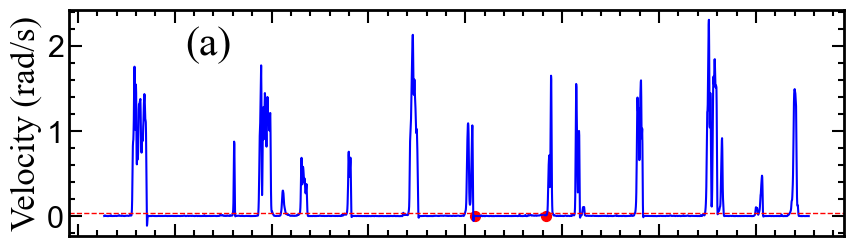} 
\includegraphics[width =.45 \textwidth]{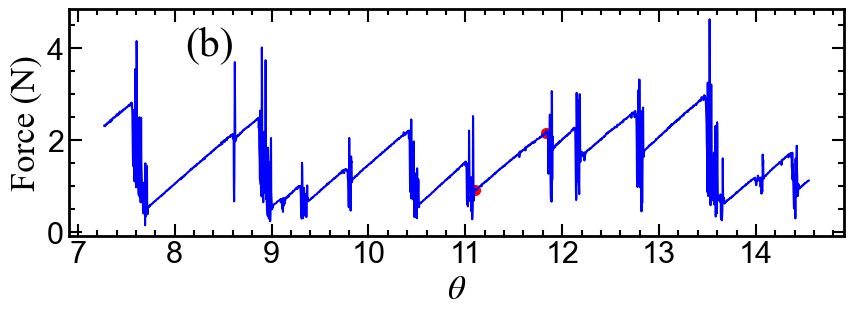}
    \caption{ Example of (a) the intruder velocity and (b) the force on the intruder showing a small part of the analyzed data set.  The full set involves hundreds of slip events. The dashed red line denotes the threshold value used to define the beginning and the end of slip events, as described in the text.  The red dots show the beginning and the end of the particular stick period which is considered in more detail in the follow-up figures. }
\label{fig:example_vel_force}
\end{figure}

Figure~\ref{fig:example_vel_force} illustrates the stick-slip dynamics of the intruder.  Part (a) shows a short sample of the intruder velocity in the azimuthal ($\theta$) direction. Note that full data sets involve several hundred stick-slip events, allowing for statistical analysis of the intruder's dynamics and of the force networks.  Figure~\ref{fig:example_vel_force}(b) shows the time-dependent force on the intruder due to loading by the torsional spring; this force is used for the normalization of the particle forces, as described in Sec.~\ref{sec:methods}.

\begin{figure}[t!]
\centering
\includegraphics[width =.45 \textwidth]{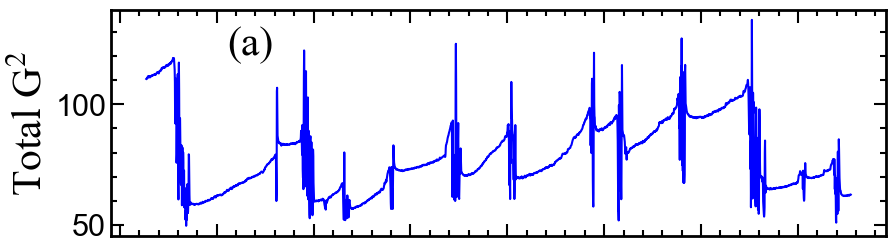} 
\includegraphics[width =.45 \textwidth]{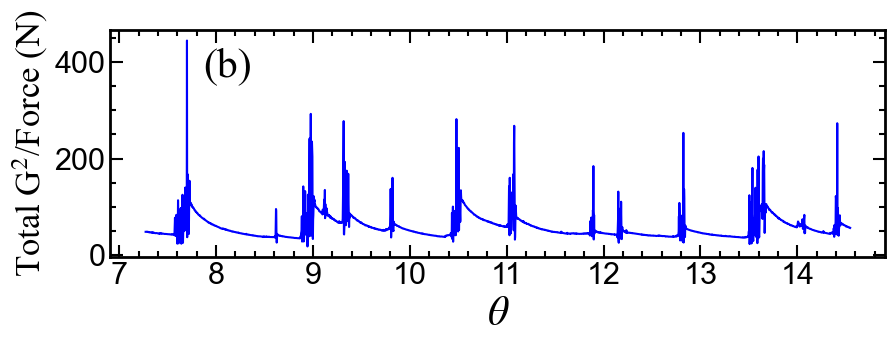}
    \caption{ G$^2$ signal for the data shown in Fig.~\ref{fig:example_vel_force}, without (a) and with (b) normalization by the intruder force shown in Fig.~\ref{fig:example_vel_force}(b).  
 }
\label{fig:example_g2}
\end{figure}

Before proceeding with the precise definition of slip events and discussion of topological measures describing the force networks, we digress briefly to comment on one aspect of the experimental results related to the normalization mentioned above that may not be obvious.  Supplemental movie 2~\cite{sup_mat1} 
has already illustrated an overall decrease in the force magnitude during sticking periods.  Figure~\ref{fig:example_g2} presents the time series for total G$^2$ (summed up over all particles), both without (a) and with (b) normalization by the intruder force.  Focusing on the stick periods, we note that the unnormalized G$^2$ signal becomes stronger, as expected due to the continuously increasing load on the intruder.  The normalized G$^2$ signal (part (b)), however, {\it decreases} during stick periods, showing that the integral of $G^2$ over all of the particles increases at a slower rate than the force on the intruder~\cite{comment}.  Further research will be needed to quantify more precisely the factors that determine which part of the force on the intruder transfers to the granular particles in the form that leads to photoelastic response, and which is balanced by the forces not captured by photoelasticity, such as frictional forces between the particles and the substrate.

In the present work, slip events are defined by using a threshold value of $v_{\rm t}= 0.04$ rad/s~\cite{kozlowski_pre_2019} for the intruder's speed  $v$ as follows: (i) a slip starts at image $i$ if $v_{i-1}< v_{\rm t}$, $v_{i}>v_{\rm t}$, and the average speed between images $i$ and $i+50$ exceeds $v_{\rm t}$; (ii) a  slip lasts until the mean speed in images $j$ to $j+2$ drops below $v_{\rm t}$, in which case $j+2$ is defined as the end of the slip.  This definition is used to avoid identifying small fluctuations in the experimental data as slips, and in particular to avoid identifying oscillations in the intruder speed, which typically occur just before slips start, as separate slip events. 

\begin{figure}[t!]
          \centering
                 \includegraphics[width=.2\textwidth]{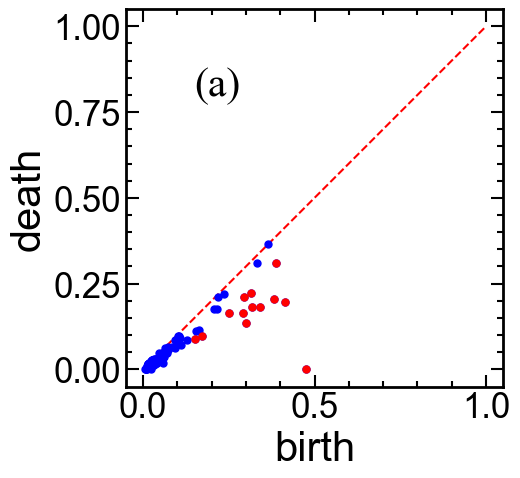}
          \includegraphics[width=.2\textwidth]{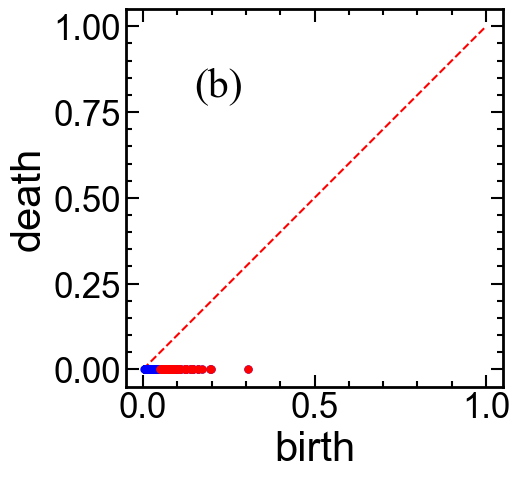}
     \caption{Persistence diagrams (PDs) corresponding to one experimental image at $\Delta \theta =0$, from the stick period denoted in Fig.~\ref{fig:example_vel_force}: (a) components, B0, (b) loops, B1.  Note that the forces used are normalized by the intruder force; blue points are inside of the `band' that is not considered due to experimental noise (see the text for a full description of how this band is defined).  Animations showing all PDs for the considered stick period are available~\cite{sup_mat4}.}
          \label{fig:b0_b1_pd_disk}
      \end{figure}

In Sec.~\ref{sec:methods} we describe how the forces on the particles and resulting force networks are defined and how the measures based on PDs are computed.  Figure~\ref{fig:b0_b1_pd_disk} shows an example of resulting PDs for components (a) and loops (b), where all forces have been normalized by the intruder force.  This figure also illustrates (in blue) the points within the band of 0.1N that are removed from further analysis, as discussed in Sec.~\ref{sec:methods}.  PDs are computed for each experimental image; animations are available~\cite{sup_mat4}.   
 
\begin{figure}[t!]
\centering
\includegraphics[width = 0.5\textwidth]{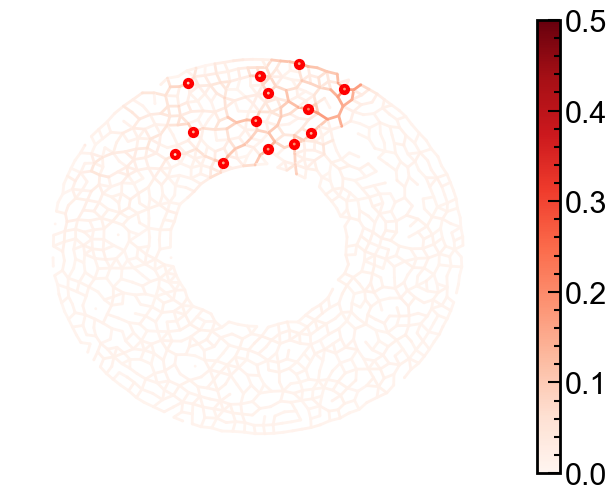}           
\caption{Generators for the B0 persistence diagram (red dots) and the corresponding force network at $\Delta \theta = 0$ for the period shown in Fig.~\ref{fig:example_vel_force}. The color bar shows the (nondimensional) force magnitude. Animation is available~\cite{sup_mat5}.   }
\label{fig:gen_networks}
\end{figure}

To illustrate the connection between the generators shown in PDs and the corresponding networks, Fig.~\ref{fig:gen_networks} shows an example of a network with particles associated with generators in PDs shown as red circles. The particles associated with the generators are identified simply by searching for particles that are assigned a force magnitude value that is within numerical precision of the birth coordinate of a generator in the PD.  This procedure is reasonably accurate, almost always allowing the identification of the relevant particles.  (For 300 figures, we identified (manually) only two cases when this procedure was not accurate in the sense that more than a single particle was connected to the same generator.)   

\begin{figure}[t!]
          \centering
         \includegraphics[width=.4\textwidth]{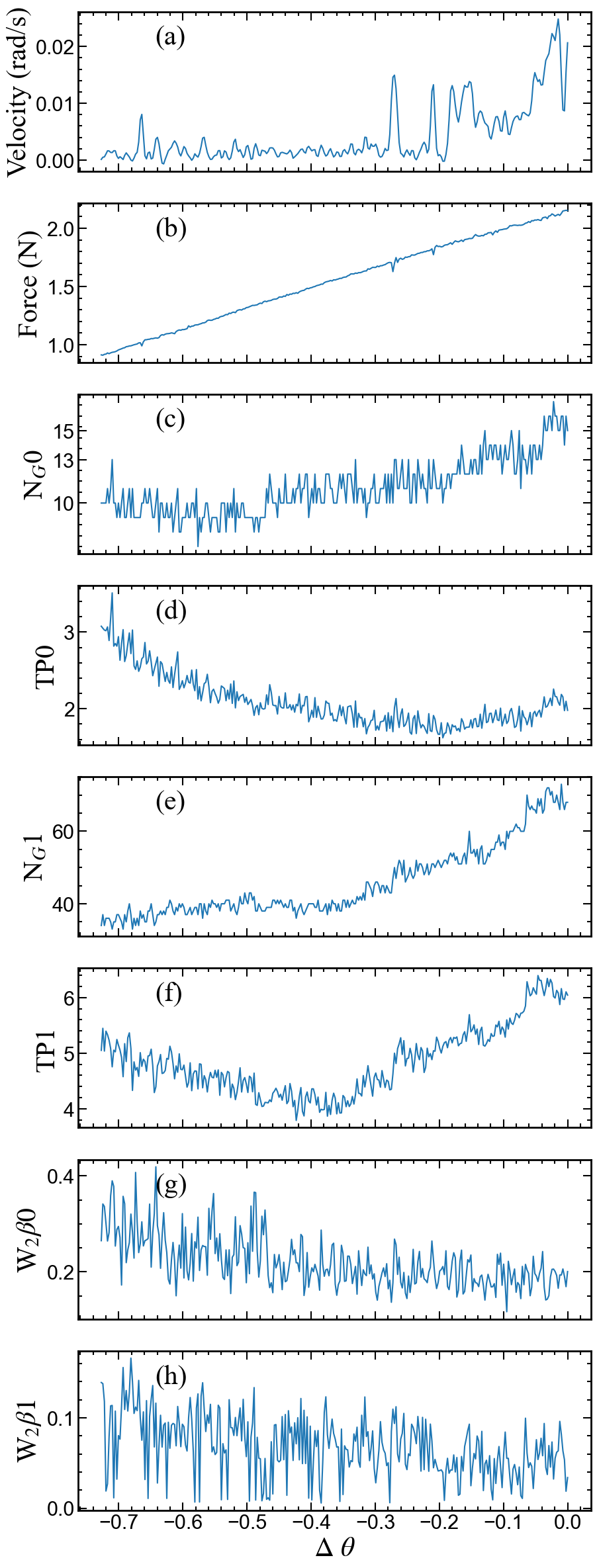}
     \caption{ Time series of the considered measures during the stick period marked by red dots in Fig.~\ref{fig:example_vel_force}: intruder's speed (a) and force (b) in the azimuthal direction,  N$_G$0 (c), TP0 (d), N$_G$1 (e), TP1 (f), W$_2\beta$0 (g), W$_2\beta$1 (h).  See the text for a description of plotted quantities. The value $\Delta \theta$ = 0 indicates the time when a slip event starts.
     }
           \label{fig:one_event_measures}
      \end{figure}

Figure~\ref{fig:one_event_measures} shows the results of persistence computations for the particular stick period denoted in Fig.~\ref{fig:example_vel_force}; animations of the resulting PDs are available~\cite{sup_mat4}.  In this figure, we introduce most of the quantities that will be considered later in Secs.~\ref{sec:disks} and \ref{sec:pentagons}: 
intruder's velocity (a) and force (b) on the intruder in the azimuthal direction, number of generators, N$_{G}$, and total persistence, TP, for components (c-d) and loops (e-f), and Wasserstein distances, W$_2 \beta 0$ and W$_2 \beta 1$ for components (g) and loops (h). Part (b) of the figure serves to remind us that, during a stick, the force on the intruder is a linearly increasing function of the drive angle; recall that this force is used to normalize all the results involving forces on the particles.  This figure also introduces some of the features of the results that will be discussed later in a more complete fashion: a slow increase of the number of generators (parts (c) and (e) of the figure), and significant activity in the measures describing dynamics of the networks, parts (g) and (h).  {\it Even though the system is in a stick period, the force networks are not static.  }

\begin{figure}[t!]
          \centering
         \includegraphics[width=.45\textwidth]{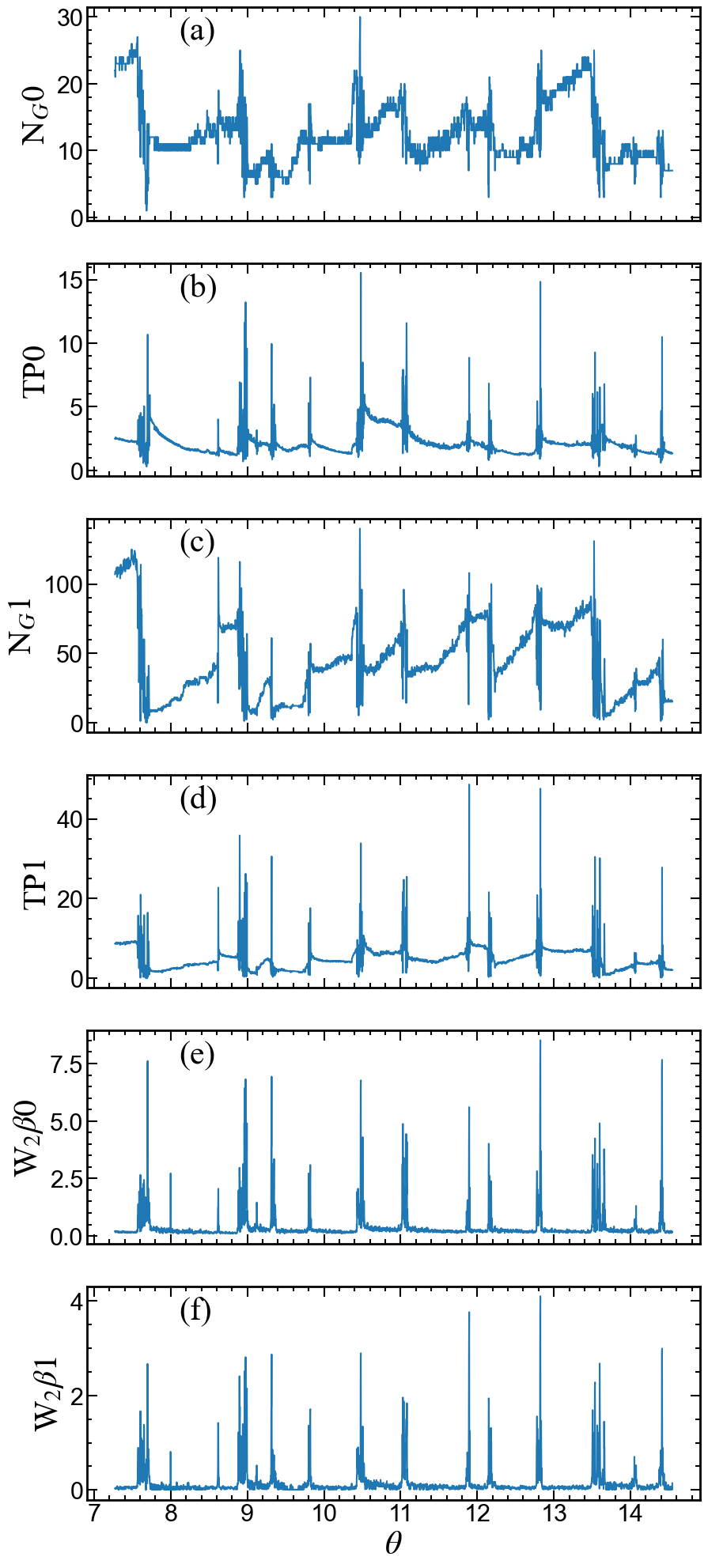}
            \caption{ Time series of the considered measures for the part of the data set shown in Fig.~\ref{fig:example_vel_force}. (a) N$_G$0, (b)TP0, (c)N$_G$1, (d) TP1, (e) W$_2\beta$0, and (f) W$_2\beta$1. }
          \label{fig:example_measures}
      \end{figure}

Figure~\ref{fig:example_measures} shows the same set of quantities plotted in Fig.~\ref{fig:one_event_measures} but now for the data set shown in Fig.~\ref{fig:example_vel_force}, therefore showing a few slip events.  We observe that the topological measures accurately follow the intruder dynamics, as expected based on the previous results obtained using simulation data~\cite{kramar2022intermittency}.  However, there is significant variability of the considered measures during the stick phases, and therefore we proceed by carrying out time-averaging of the results over a large number of stick periods.  The results of this analysis are discussed next.
      
\subsection{Disks: Full data sets}
\label{sec:disks}

\begin{figure}[t!]
         \centering
         \includegraphics[width=.45\textwidth]{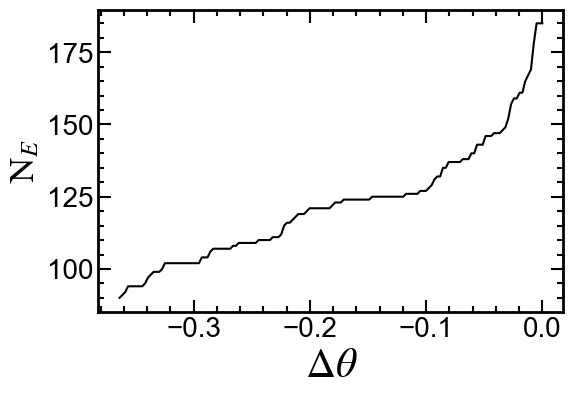}
         \caption{Number of stick periods, N$_E$, as a function of the angular distance, $\Delta \theta$, to the upcoming slip event (occurring at $\Delta \theta = 0$).
        }
         \label{fig:events_disk}
\end{figure}

The full data set for a given volume fraction includes hundreds of stick-slip events.  To analyze this data, we start by considering the behavior of various measures during a stick period as a function of the time to the next slip event (quantified here by the angle, $\Delta \theta$, of the drive shaft, where $\Delta \theta = 0$ corresponds to the upcoming slip).   Figure~\ref{fig:events_disk} shows the number of stick periods with a duration greater than  $\Delta \theta$; clearly, the number of stick periods decreases for larger (in absolute value terms) $\Delta \theta$, since some of the stick periods are shorter than the $\Delta \theta$ range considered.   We note that there is a significant number of rather short stick periods; we have carried out the analysis, discussed in what follows, by both including and excluding these short periods, with similar outcomes.  For brevity, here we will only discuss the results obtained by including all stick periods, even if very short. 
      
\begin{figure}[t!]
    \centering
    \includegraphics[width = .45\textwidth]{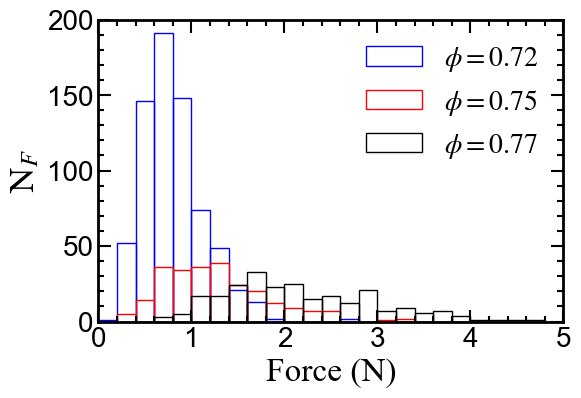}
    \caption{Distribution of the force on the intruder at which slips occur for three available volume fractions. }
    \label{fig:force_distribution_disks}
\end{figure}

Figure~\ref{fig:force_distribution_disks} shows the distribution of spring forces at which slips happen for three available volume fractions of $\phi = 0.72,~0.75$ and $0.77$.  We observe wide distributions, in particular for $\phi = 0.77$, suggesting that the spring force is not a good predictor of time of upcoming slip, as expected.

\begin{figure}[t!]
          \centering
        
         \includegraphics[width=.2\textwidth]{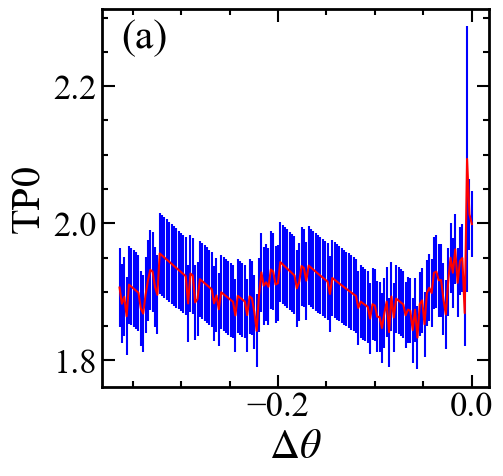}
         \includegraphics[width=.2\textwidth]{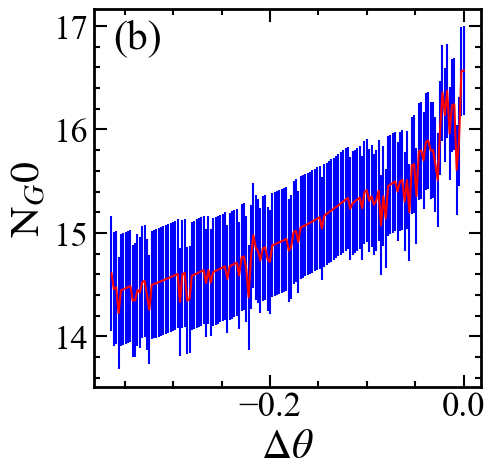} 
         \includegraphics[width=.2\textwidth]{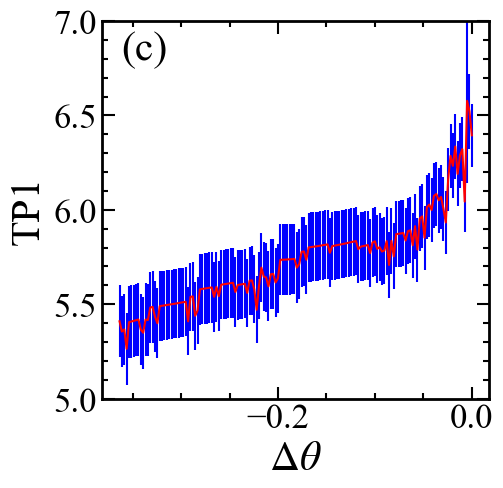}
         \includegraphics[width=.2\textwidth]{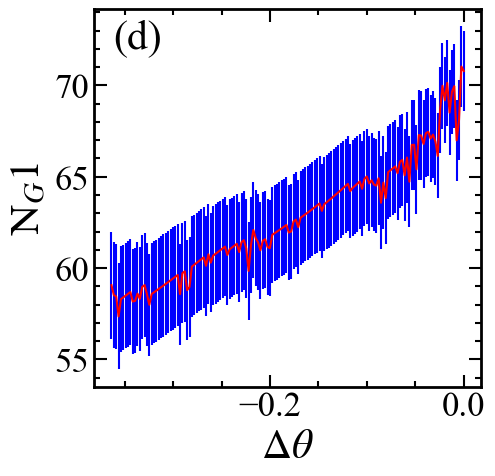}
         \includegraphics[width=.2\textwidth]{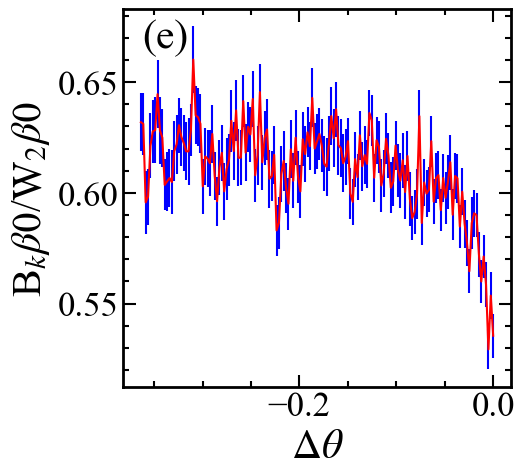}
         \includegraphics[width=.2\textwidth]{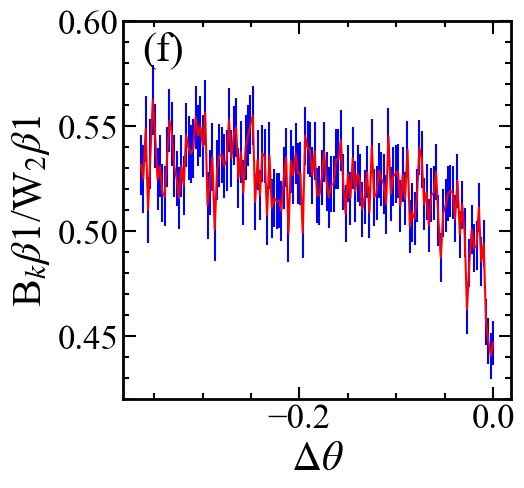}
         
     \caption{Persistence-derived measures averaged over all stick periods (a-b) TP and the number of generators for components; (c-d) TP and generators for loops; (e-f) ratio of B$_k$ to W$_2$ for components and loops. Red lines show the averages, and blue lines their standard error. } 
          \label{fig:disks_tp_gen_av}
      \end{figure}

We proceed by discussing averages over all stick periods as a function of $\Delta\theta$, focusing on $\phi = 0.77$ only.   Figure~\ref{fig:disks_tp_gen_av} shows the results.  
To help their interpretation, it is important to remember that the forces used for computing PDs and resulting measures are normalized by the current value of the intruder force; therefore a constant mean value of the shown quantities would suggest that the force network simply scales linearly with the intruder force.  This is however not what we observe, and therefore we conclude that the force networks evolve in a non-trivial way even during stick periods when particles are essentially stationary.  Note that we include only the experimental data from the times when the force on the intruder is larger than 0.1N; this restriction prevents normalizing by a very small intruder force that may occur during slips.  

The main finding illustrated by Fig.~\ref{fig:disks_tp_gen_av}(b,~d) is a continuous monotonic increase of the number of generators, N$_G$, both for components and for loops. This finding indicates an increase in the complexity of the considered networks as the upcoming slip is approached: to use the landscape analogy mentioned in Sec.~\ref{sec:methods}, the number of peaks in the force landscape increases. This trend is not influenced by the width of the band of generators close to the diagonal that is excluded: we have verified that when the width of the band is halved, the trend remains the same.  The trends for TP are less clear than the ones for the number of generators.  In the case of components, panel~(a), we see an essentially constant mean value, suggesting that the force networks become smoother as the slip is approached. (Since the number of generators grows, constant TP implies that on average the generators move closer to the diagonal. Using the landscape analogy, the average height of the peaks in the force network decreases.)  Regarding loops, part (c), the fact that the number of generators and TP grow at a similar relative rate suggests that at least on average the strength of the normalized forces leading to loops remains constant. We have confirmed that smaller volume fractions ($\phi = 0.75$ and 0.72) lead to consistent results (figures not shown for brevity).  

Figure~\ref{fig:disks_tp_gen_av}(e-f) provides a different measure characterizing force networks. Here we plot the ratio of two distances (discussed in Sec.~\ref{sec:methods}) that describe the changes in the force network from one image to the next.  However, B$_k\beta_0$ measures the largest distance only, while W$_2\beta_0$ is the square root of the  sum of  the squares of all  distances in the matching.  If there is a single  significant change, then this ratio is close to unity.  If the ratio is much smaller than unity, that means that there are differences between the networks on a scale comparable to the largest one.   From Fig.~\ref{fig:disks_tp_gen_av}(e~-~f) we see that the ratio for both components is in the range [0.60-0.65] for loops in the range [0.50-0.55] sufficiently far from the upcoming slip.  This means that for most of the stick duration, there is a single change in the network that carries a lot of weight; closer to the upcoming slip, however, the ratio decreases, showing that closer to the slip there are more changes in the force network, and therefore the dominance of one single change weakens, as expected based on the previous works exploring slip precursors~\cite{Talamali2011_pre,scuderi_natgeoscience_2016,lherminier_PRL_2019,johnson_geophys_2013,johnson_grl_2017,nerone_pre_2003,aguirre_pre_2006,scheller2006precursors,zaitsev_epl_2008,staron2002preavalanche,Maloney2004_prl,ciamarra_prl10,welker2011precursors,dahmen_natcom_2016,dahmen_group15,Denisov2017_sr,bares_pre17}.      

\begin{figure}[t!]
    \includegraphics[width = .2\textwidth]{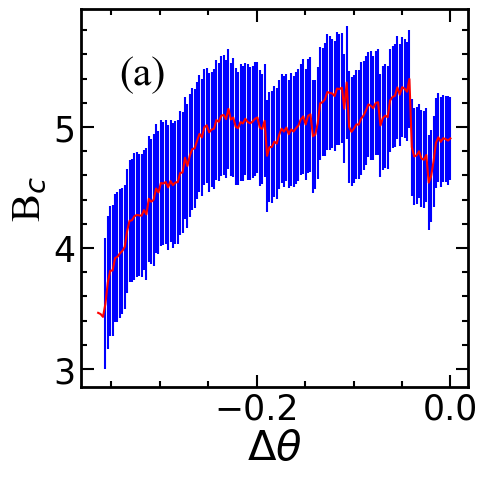}
    \includegraphics[width = .22\textwidth]{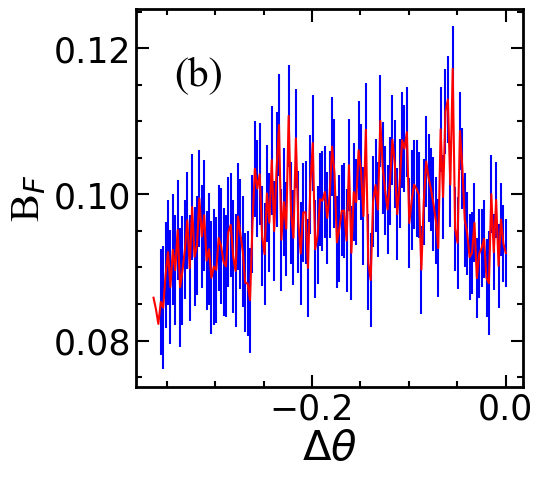}
    \caption{Number of broken contacts and average broken force (averaged over all stick periods). }
    \label{fig:bc}
\end{figure}

We next discuss whether the precursors mentioned above could be related to broken contacts between the particles.  Figure~\ref{fig:bc}(a)  plots the number of broken contacts,  B$_c$, defined as the number of contacts that break between consecutive images (without accounting for newly formed contacts), in the same format as used in  
 Fig.~\ref{fig:disks_tp_gen_av}.  For completeness, Fig.~\ref{fig:bc}(b) plots also the average broken force. The main observation is that neither B$_c$, nor the `broken force' magnitude, increase close to the upcoming slips, at least on average.  We conclude that the precursors to slip may develop even without breaking physical contacts between the particles.

We have also explored whether there is a correlation between N$_G$ and/or TP and the quantities such as stick duration, or consequent slip size, but were not able to extract any meaningful results (figures not shown for brevity). Further work will be needed to clarify the generality and consequences of this finding.

\begin{figure}[t!]
         \centering
             \includegraphics[width=.22\textwidth]{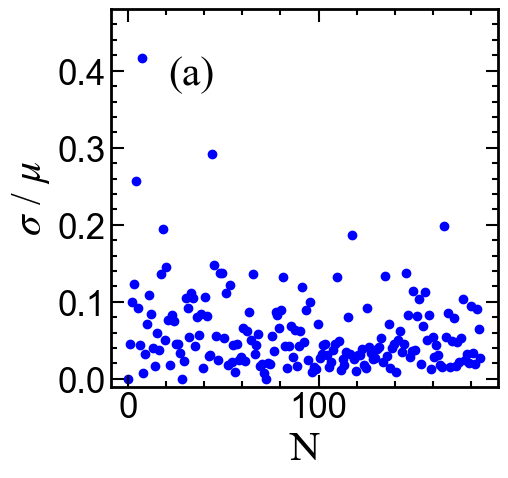}
             \includegraphics[width=.2\textwidth]{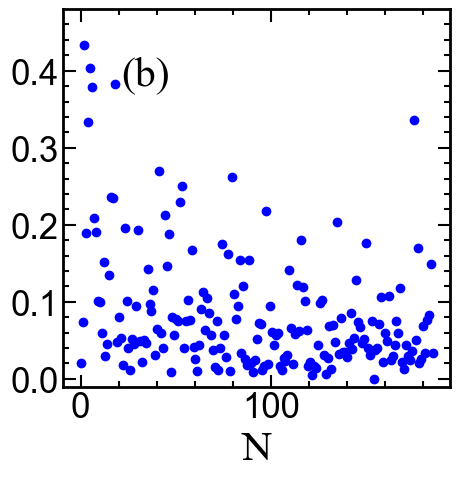}

     \caption{The ratio of the standard deviation ($\sigma$) to the average ($\mu$) of (a) N$_G$0, (b)N$_G$1 for disks. N counts stick events; each point in the plot corresponds to a separate stick event. 
       }
          \label{fig:ratio_sigma_mu_each_event}
      \end{figure}

Figure~\ref{fig:ratio_sigma_mu_each_event} shows the ratio of the standard deviation to the average number of generators during each stick period. More precisely, for each stick period, we compute the mean number of generators, $\mu$, and the standard deviation, $\sigma$, by using the information from all the images corresponding to a stick period.  Therefore, here we focus on variability during each stick, in contrast to the results shown in Fig.~\ref{fig:disks_tp_gen_av}, which focus on the average over many sticks. The $\sigma/\mu$  ratio is reasonably small ($< 0.1$) for most of the stick periods, showing that, while the number of generators varies considerably between stick periods, during any given stick period, the number of generators (and therefore at least some features of the force network itself) do not change much. We will see below that this finding does not apply to pentagons.  

\subsection{Examples of time-dependent results for pentagons}
\label{sec:time_dep_pentagons}

\begin{figure}[t!]
\centering
\includegraphics[width = .45\textwidth]{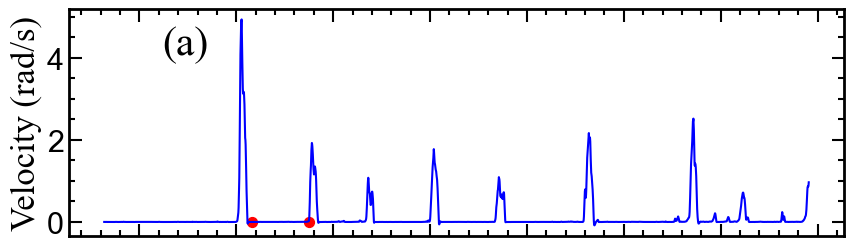} 
 
\includegraphics[width =.45 \textwidth]{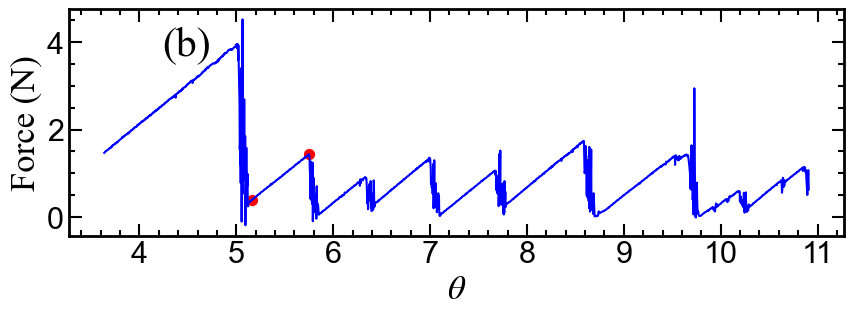}

    \caption{ Example of (a) intruder's velocity and (b) force on the intruder showing a small part of the analyzed data set for pentagons, $\phi = 0.65$. The red dots show the beginning and the end of a particular stick period that we consider in more detail in the follow-up figures. }
\label{fig:example_vel_force_pent}
\end{figure}

In this section, we show examples of the results obtained using pentagon-shaped particles, focusing on the features that are different from those obtained for disks.  Figure~\ref{fig:example_vel_force_pent} illustrates the stick-slip dynamics of the intruder using a small portion of the full pentagon data set for $\phi = 0.65$  (the largest value with which the present experiments could be carried out).  We again mark by red dots the beginning and end of a particular stick period that will be considered in more detail next.  

\begin{figure}[t!]
          \centering
    
         \includegraphics[width=.2\textwidth]{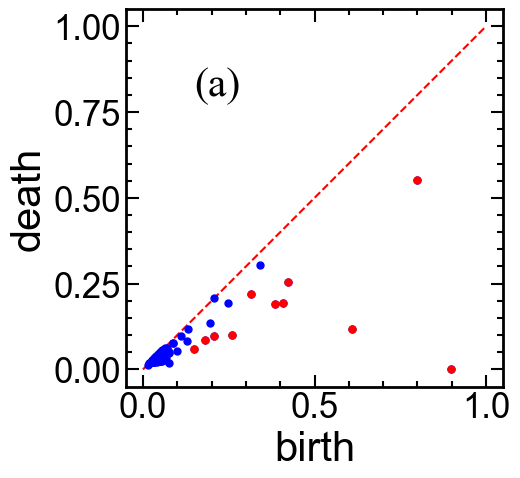}
         \includegraphics[width=.2\textwidth]{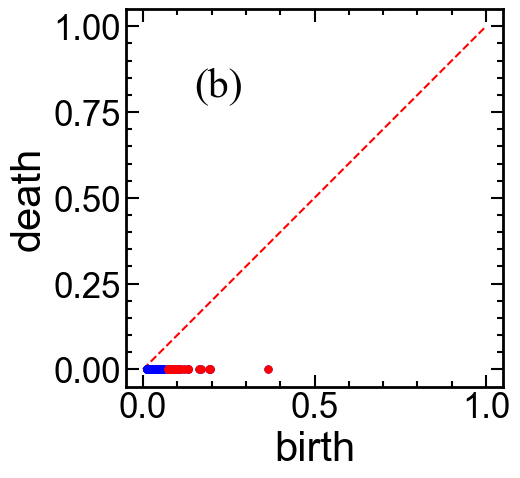}
         
     \caption{Persistence diagrams (PDs) corresponding to the experimental image at $\Delta \theta=0$ from the stick period denoted in Fig.~\ref{fig:example_vel_force_pent} (animations of all PDs corresponding to this stick period are available~\cite{sup_mat6}.  (a) components, B0, (b) loops, B1.   }
          \label{fig:b0_b1_pd_pent}
      \end{figure}

\begin{figure}[thbp!]
\centering
\includegraphics[width = 0.5\textwidth]{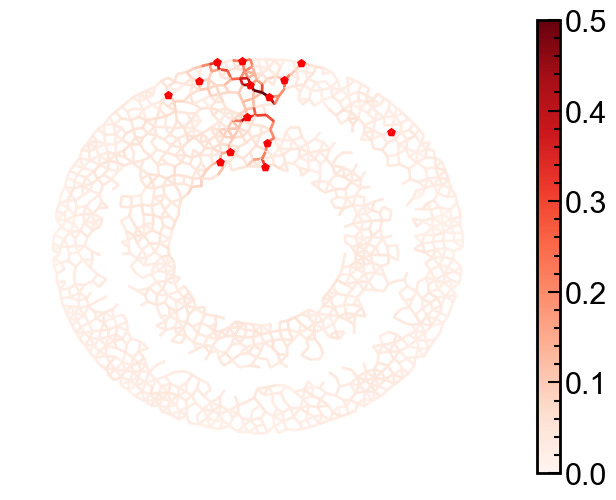}           
\caption{Generators for a PD and corresponding network at $\Delta \theta = 0$ for the period shown in Fig.~\ref{fig:example_vel_force_pent}. The color bar shows the (nondimensional) force magnitude. Animation is available~\cite{sup_mat7}.}
\label{fig:gen_networks_pent}
\end{figure}

Figure~\ref{fig:b0_b1_pd_pent} shows sample PDs for the considered stick (animations are available~\cite{sup_mat6}).  The computed networks with superimposed particles corresponding to generators in the PDs are shown in Fig.~\ref{fig:gen_networks_pent}.  This figure illustrates one significant difference between disks and pentagons: for values of $\phi$ that result in stick-slip dynamics, pentagon-based systems form an open channel (except immediately in front of the intruder). As the intruder advances, the pentagons in front of it support forces that resist the intruder's motion and move aside when the system slips. For disks, on the other hand, packing densities that yield similarly long open channels allow for continuous motion of the intruder with only occasional clogging events; stick-slip motion is observed only at higher $\phi$ where the open channel behind the intruder is rather short, as can be seen in Fig.~\ref{fig:gen_networks}(a). We note that the formation of the open channel has been discussed in the literature already, and the reader is directed to Refs.~\cite{kozlowski_pentagonsanddisks2021,carvalho2022,kolb2013,seguin2016} and references therein for further discussion.  We emphasize that the configuration shown in Fig.~\ref{fig:gen_networks_pent} occurs during the steady-state part of the intruder's dynamics: by the time shown in the figure, the intruder has already traveled more than four revolutions around the annulus.

The time series of the results for pentagons are similar to the ones for disks, so we avoid showing corresponding figures for brevity. The main observation is that all considered measures shown in Figs.~\ref{fig:one_event_measures} and~\ref{fig:example_measures} can be computed for pentagons despite the different particle shape. We expect that this finding is a consequence of the stability of persistence diagrams to small perturbations, once the contact network is known.

\subsection{Pentagons: Full data sets}
\label{sec:pentagons}      
      \begin{figure}[t!]
         \centering
         \includegraphics[width=.45\textwidth]{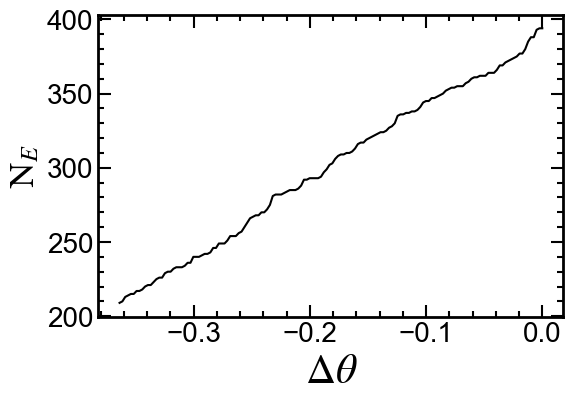}
         \caption{Pentagons: N$_E$ as a function of $\Delta \theta$. }

         \label{fig:events_pent}
     \end{figure}

\begin{figure}[t!]
    \centering
    \includegraphics[width = .45\textwidth]{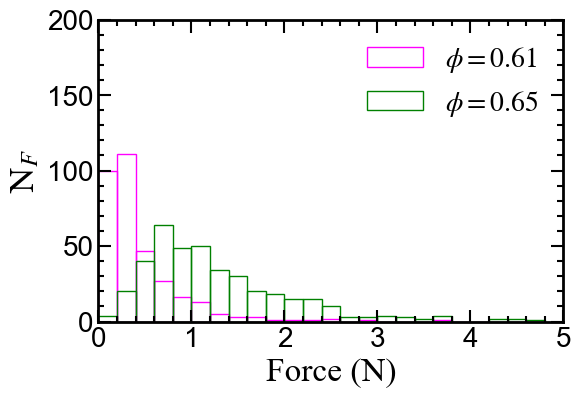}
    \caption{Distribution of the force on the intruder at which slip occurs for different volume fractions.}
    \label{fig:force_distribution_pent}
\end{figure}

Figure~\ref{fig:events_pent} shows the number of stick periods longer than a given duration, with similar behavior as the one seen for the disks.  Figure~\ref{fig:force_distribution_pent} shows the distribution of the force on the intruder for pentagons at which slips occur for $\phi = 0.65$ and $0.61$.  Similarly as for disks, the peak of the distribution moves to a lower value of force as $\phi$ is decreased. Again, the spread of values prevents the use of force on the intruder as a precursor for the upcoming slip, as is usual in stick-slip dynamics. 
      
\begin{figure}[t!]
          \centering
          
         \includegraphics[width=.2\textwidth]{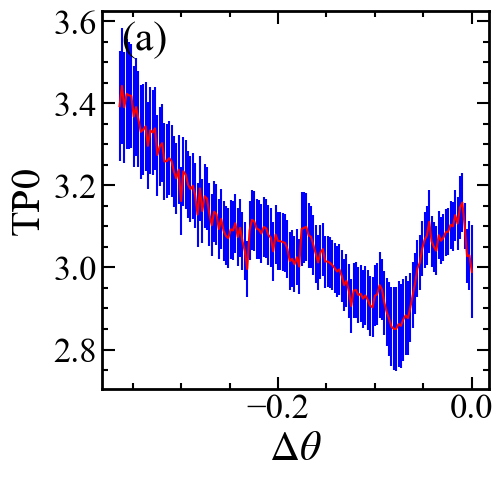}
         \includegraphics[width=.2\textwidth]{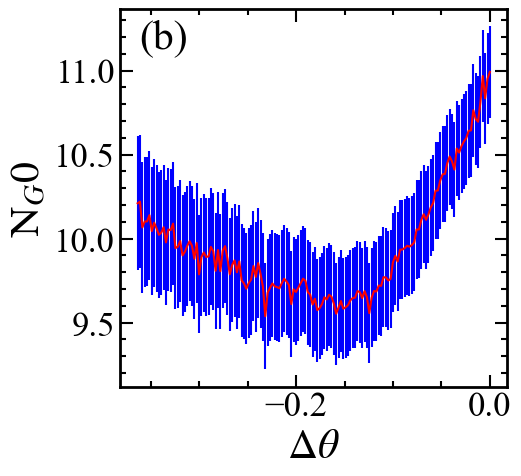}
         \includegraphics[width=.2\textwidth]{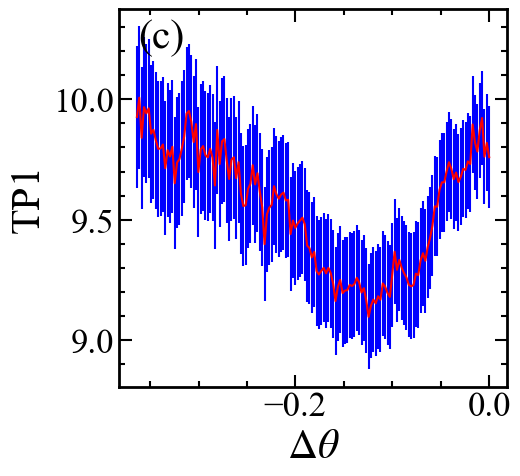} 
         \includegraphics[width=.2\textwidth]{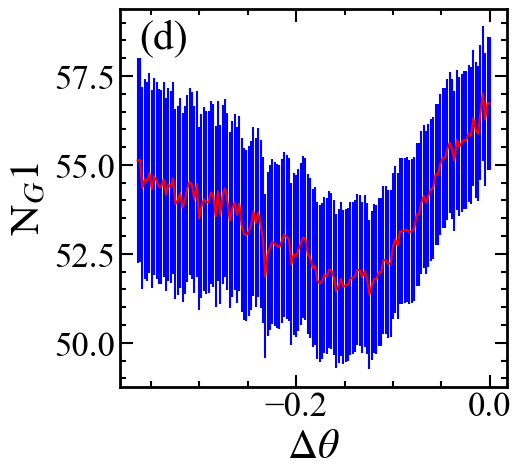}
         \includegraphics[width=.2\textwidth]{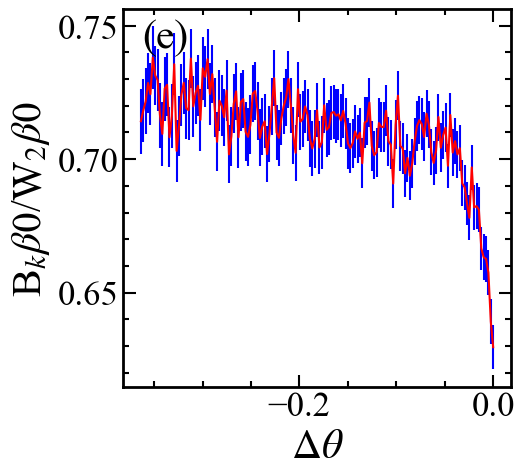} 
         \includegraphics[width=.2\textwidth]{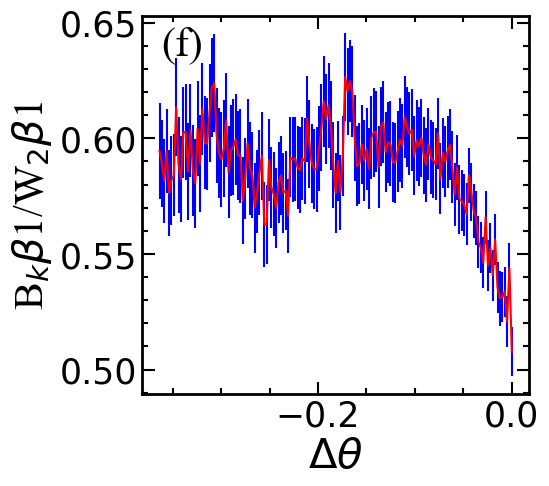} 
         
     \caption{Persistence-derived measures averaged over all stick periods: (a-b) TP and number of generators for components; (c-d) TP and number of generators for loops; (e-f) 
     ratio of B$_k$ to W$_2$ for components and loops.}
          \label{fig:avg_pent}
      \end{figure}

We proceed by briefly discussing the measures averaged over many sticking periods.  Figure~\ref{fig:avg_pent} shows the results for total persistence, the number of generators, and the ratio of distances (vis. similar results for disks, Fig.~\ref{fig:disks_tp_gen_av}). One difference between disks and pentagons can be seen in the behavior of the number of generators, N$_G$: while for disks we always observe a monotonic increase, for pentagons an increase is found only close to the upcoming slips (for small absolute values of $\Delta \theta$).   The results for the ratio of distances, Fig.~\ref{fig:avg_pent}(e-f) are similar to those for disks, suggesting once again the presence of slip precursors.

\begin{figure}[t!]
             \includegraphics[width=.24\textwidth]{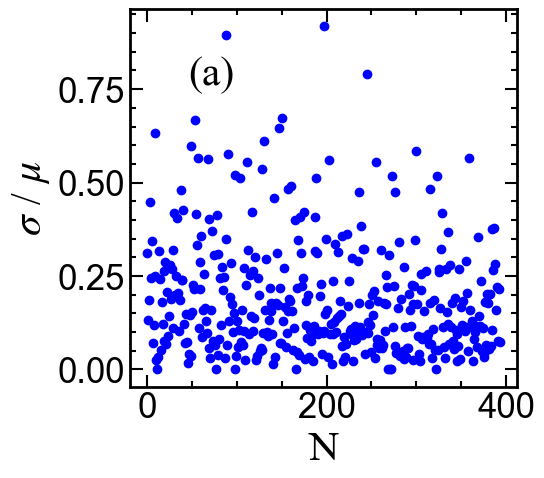}
             \includegraphics[width=.2\textwidth]{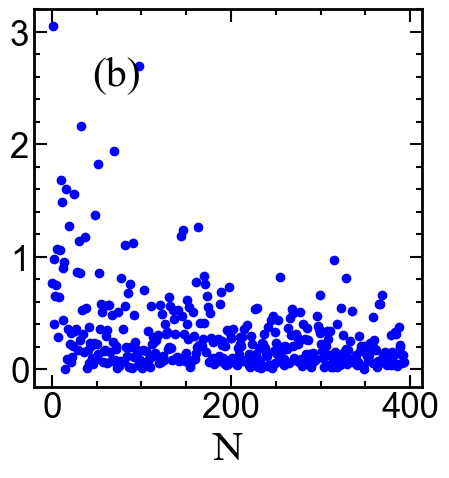}

     \caption{The ratio of the standard deviation ($\sigma$) to the average ($\mu$) (a) N$_G$0, (b)N$_G$1 for pentagons (viz. Fig.~\ref{fig:ratio_sigma_mu_each_event} for disks).
          }
          \label{fig:pentagon_ratio_sigma_mu_each_event}
      \end{figure}

Figure~\ref{fig:pentagon_ratio_sigma_mu_each_event} illustrates yet another difference between pentagons and disks: here we plot the ratio of the standard deviation to the average value of the number of generators.  Comparing with  Fig.~\ref{fig:ratio_sigma_mu_each_event}, we find that the variability of the number of generators is significantly larger for pentagons.  This indicates that during a typical stick, the networks vary much more for pentagons than for disks. Such an observation can be easily obtained using persistence analysis, while it may be difficult to reach by considering the force networks directly.  

\section{Conclusions}
\label{sec:conclusions}

The main goal of the present paper is to illustrate an approach to analyzing stress networks in granular systems that does not require the resolution of individual contact forces. Networks based on the G$^2$ averaged over a particle (known to describe accurately the sum of the magnitudes of the particle's normal contact forces) are readily obtained from experimental images and provide useful information when analyzed using tools of persistent homology.  We hope that the approach outlined here will be useful to the researchers exploring a number of other granular systems for which the sum of the magnitudes of the normal contact forces is available either via photoelasticity or some other means.   

We find that the quantification of the considered networks leads to measures that precisely identify stick and slip regimes of an intruder traveling in a medium containing photoelastic particles. The implemented measures show that force networks evolve even when the dynamics is essentially absent and the intruder is stationary. This conclusion applies even when the considered force network is scaled by the current value of the intruder's force. 
Furthermore, the evolution in the stick phase typically involves a gradual increase in the number of generators in persistence diagrams, corresponding roughly to the formation of more elaborate networks with an increased number of components and loops.  We find that the changes of the networks during a single stick period are significantly smaller for disks compared to pentagons. More generally, however, we do not find any dramatic differences between stick-slip dynamics and force network properties for disks and pentagons.

The considered dynamic measures, based on computing the distances between consecutive persistence diagrams, show that as a slip approaches, there are more and more force network changes.  While this finding is not necessarily surprising (precursors to slip events have been discussed extensively in the literature already), we find it encouraging to see that such information could be extracted directly from consideration of the force networks.  We, therefore, hope that further analysis of force networks for systems experiencing intermittent dynamics will find its role in the continuing quest for novel approaches to the prediction of large events.  

This work was supported by the U.S. Army Research Office through Grant No. W911NF-18-1-0184 and by the Keck Foundation. R.K. thanks the Department of Physics, Duke University, where the experimental work was conducted. M.C. is thankful for the financial support from CONICET (Grant No. PUE 2018 229 20180100010 CO) and UTN (Grant No. MAUTILP0007746TC).


\begin{thebibliography}{55}
\bibitem [{com()}]{comment}%
  \BibitemOpen
  \href@noop {} {}\bibinfo {note} {Careful reader will recall the cubical fit
  used for calibration of G$^2$ signal; we have confirmed that the finding
  regarding a decrease of unscaled G$^2$ translates also to a decrease of the
  force itself.}\BibitemShut {Stop}%
\end{thebibliography}

%

\end{document}